\documentclass[aps,showpacs,prd,twocolumn]{revtex4}
\usepackage{amssymb}
\usepackage{amsfonts}
\usepackage{amsmath}
\usepackage{graphicx}
\usepackage{color}
\usepackage[dvips]{epsfig}
\usepackage[dvips]{graphicx}
\usepackage{float}

\newcommand{\be}{\begin{equation}}
\newcommand{\ee}{\end{equation}}
\newcommand{\bes}{\begin{subequations}}
\newcommand{\ees}{\end{subequations}}
\newcommand{\ben}{\begin{eqnarray}}
\newcommand{\een}{\end{eqnarray}}

\begin{document}

\title{Trapping Dirac fermions in tubes generated by two scalar fields}
\author{R. Casana$^{1}$, A. R. Gomes$^{2}$, G. V. Martins$^{1}$, F. C. Simas$%
^{1}$}
\affiliation{$^{1}$Departamento de F\'{\i}sica, Universidade Federal do Maranh\~{a}o,\\
65085-580, S\~{a}o Lu\'{\i}s, Maranh\~{a}o, Brazil.}
\affiliation{$^2$Departamento de F\'isica, Instituto Federal do Maranh\~ao,\\
65025-001, S\~ao Lu\'\i s, Maranh\~ao, Brasil}

\begin{abstract}
In this work we consider $(1,1)-$dimensional resonant Dirac fermionic states
on tube-like topological defects. The defects are formed by rings in $(2,1)$
dimensions, constructed with two scalar field $\phi$ and $\chi$, and
embedded in the $(3,1)-$dimensional Minkowski spacetime. The tube-like
defects are attained from a lagrangian density explicitly dependent with the
radial distance $r$ relative to the ring axis and the radius and thickness
of the its cross-section are related to the energy density. For our purposes
we analyze a general Yukawa-like coupling between the topological defect and
the fermionic field $\eta F(\phi,\chi)\bar\psi\psi$. With a convenient
decomposition of the fermionic fields in left- and right- chiralities, we
establish a coupled set of first order differential equations for the
amplitudes of the left- and right- components of the Dirac field. After
decoupling and decomposing the amplitudes in polar coordinates, the
radial modes satisfy Schr\"odinger-like equations whose eigenvalues are the
masses of the fermionic resonances. With $F(\phi,\chi)=\phi\chi$ the
Schr\"odinger-like equations are numerically solved with appropriated
boundary conditions. Several resonance peaks for both chiralities are
obtained, and the results are confronted with the qualitative analysis of
the Schr\"odinger-like potentials.
\end{abstract}

\pacs{11.10.Lm,11.27.+d}
\maketitle


\section{introduction}

Braneworld scenarios have their origins in attempts of solution of important
problems of theoretical physics such as the cosmological constant and the
gauge hierarchy \cite{br1,br2,br3,br4,br5}. In the original formulation of
thin branes, the matter fields are by construction localized on a brane with
energy density described by a delta function \cite{thin}, while gravity
propagates in all dimensions. Usual Newton's law can then be reproduced on
the brane depending on the metric warp factor, attained after solving
Einstein's equations. Several extensions  soon appeared, with
smooth thick branes constructed by scalar fields \cite%
{thick1,thick2,thick3,thick4,thick5,thick6,thick7,thick8,thick9,thick10}. A
comprehensive review on this subject can be found in \cite{thick_rev}. This opened up the
idea of matter fields to visit extra dimension space, with possible signal
of deviations of standard model due to extra dimensions.

In general, thick branes are possibly able to trap gravitons and scalar
fields. For fermions, however, the introduction of the fermion-scalar
coupling is a necessary condition to ensure the normalizable zero modes.
This is a known property already demonstrated by Jackiw and Rebbi \cite%
{jackiw} for domain walls. For some models, the massive fermionic states
leak from the branes but stay for a sufficiently longer time to be
characterized as resonances \cite%
{brss1,brss2,brss3,brss4,brss5,brss6,brss7,brss8,brss9}. In particular, the
Ref. \cite{liu} analyzes the localization of matter fields in branes
constructed from a scalar field coupled to a dilaton. In \cite{liu1}, the localization and
mass spectra of various matter fields in thick $AdS$ brane were
investigated. For fermionic Kaluza-Klein modes, bound states for both
chiralities were found. In \cite{brss5}, some of the authors of the present
work have investigated the presence of massive modes for right-hand and
left-hand fermions with branes with internal structure constructed by two
scalar fields coupled to gravity by introducing a simple Yukawa coupling.

In this work we are interested in topological defects embedded in a flat
spacetime that can be constructed following similar procedure used for
modeling branes, namely, the embedding of a topological defect in one or
more extra dimensions. Thus, inspired by the physics of extra dimensions,
in Section II we consider $(2,1)$-dimensional ring-like topological defects
\cite{bmm1,bmm2} which are embedded in a $(3,1)$-dimensional flat
spacetime forming a tube-like topological defect. In Section III, we study
some aspects of localization of fermionic fields in this system. We have
particular interest for resonance effects, which are studied in Section IV. Our
conclusions are presented in Section V.

\section{A tube in $(3,1)$-dimensions}

A tube in $(3,1)$-dimensions can be described by the action
\begin{equation}
S_{tube}=\!\int\!\! dtd^{3}x \left(\frac{1}{2}\partial _{M}\phi \partial
^{M}\phi +\frac{1}{2}\partial _{M}\chi \partial ^{M}\chi -V(\phi ,\chi
)\right) ,  \label{actionring}
\end{equation}%
with
\begin{equation}
V(\phi ,\chi )=\frac{1}{2r^{2}}(W_{\phi }^{2}+W_{\chi }^{2}).
\end{equation}%
We use capital letters $M$, $N$ for all $(3,1)$ dimensions. The explicit
dependence of $r=\sqrt{y^{2}+z^{2}}$ follows closely and generalizes for two
fields the construction of \cite{bmm1,bmm2} for evading Derrick-Hobarts'
theorem \cite{raj1,raj2,raj3}.

We notice that this construction breaks translational invariance, which is
also present in QCD scenarios. For instance, in investigations which deal
with color superconductivity, pairing with quarks with different chemical
potentials results in crystalline quark matter condensates which
spontaneously break translational and rotational invariance, and include
spin-zero Cooper pairs \cite{color1a,color1b}. In \cite{color2a,color2b} the effective
Lagrangian density describing the color-flavor locked (CFL) symmetry phase
of QCD at high density has fields depending on the velocity of the massless
Dirac fermions. With glueball effective lagrangian model the breaking of
Lorentz invariance induced by the quark chemical potential affect the
critical temperature for the onset of the superconductive state \cite{deconf}%
. The breaking of translational invariance also occurs in problems dealing
with brane intersections \cite{br-inta,br-intb}, noncommutative field theory with
non-constant noncommutativity \cite{n-coma,n-comb} and condensed matter physics \cite%
{dobr,bor}.

The equations of motion for static solutions are
\begin{eqnarray}
\frac{1}{r}\frac{d}{dr}\bigg(r\frac{d\phi }{dr}\bigg) &=&\frac{1}{r^{2}}%
(W_{\phi }W_{\phi \phi }+W_{\chi }W_{\chi \phi })  \label{eqmotionphi} \\
\frac{1}{r}\frac{d}{dr}\bigg(r\frac{d\chi }{dr}\bigg) &=&\frac{1}{r^{2}}%
(W_{\phi }W_{\phi \chi }+W_{\chi }W_{\chi \chi }) .  \label{eqmotionchi}
\end{eqnarray}%
In this work we restrict us to configurations with radial symmetry, i.e.,
the fields $\phi =\phi (r)$ and $\chi =\chi (r)$ depends only on $r$. One
can show that the solutions of the first-order equations \cite{bmm1}
\begin{equation}
\frac{d\phi }{dr}=\frac{1}{r}W_{\phi }~\ ,~\ \frac{d\chi }{dr}=\frac{1}{r}%
W_{\chi }  \label{eqfirstorderchi}
\end{equation}%
are also solutions of the second-order equations (\ref{eqmotionphi}) and (%
\ref{eqmotionchi}). The change of variables $d\xi =(1/r)dr$ effectively
turns the $2$-dimensional model in a one-dimensional one, since Eqs. (\ref%
{eqfirstorderchi}) can be rewritten as
\begin{equation}
\frac{d\phi }{d\xi }=W_{\phi }~\ ,~\ \frac{d\chi }{d\xi }=W_{\chi }.
\label{eqfirst1Dchi}
\end{equation}

In this work, for generating the tube solution we consider \cite{bsr}
\begin{equation}
W(\phi ,\chi )=\lambda \left( \phi -\frac{1}{3}\phi ^{3}-s\phi \chi
^{2}\right)  \label{superpotential}
\end{equation}%
This choice of $W(\phi,\chi)$ with the potential $\tilde{V}(\phi ,\chi
)=(1/2)(W_{\phi }^{2}+W_{\chi }^{2})$ was studied in Ref. \cite{bsr}. The
potential $\tilde{V}(\phi ,\chi )$ has minima at $(\pm 1,0)$ and $(0,\pm
\sqrt{1/s})$ with $s>0$ and the equations of motion has static solutions
connecting the minima $(\pm 1,0)$ as defects with internal structure known
as Bloch walls. The limit $s\rightarrow 0.5$ turns the two-field problem
into a one-field model with solution known as Ising wall. See also Refs.
\cite{shif, izq} for other solutions. The extension of this construction to $%
(4,1)$ dimensions leading to Bloch brane was presented in \cite{thick9}.  The richer structure of degenerate and critical Bloch branes
where proposed in \cite{dah}. In \cite{thick9} it was shown that the presence of the field $\chi $ is crucial to
giving internal structure to the brane. In the present work the presence of
the field $\chi $ also contributes to generate an internal structure to the
tube formed. We will see that this is crucial for localizing fermions with
a simple Yukawa coupling.

The choice given by Eq. (\ref{superpotential}) generates the known $1 $%
-dimensional solutions for Eqs. (\ref{eqfirst1Dchi})
\begin{eqnarray}
\phi (\xi ) &=&\,\mathrm{tanh}\big(2\lambda s\xi \big),  \notag \\[-0.2cm]
&&  \label{solringchi} \\[-0.2cm]
\chi (\xi ) &=&\pm \sqrt{\frac{1}{s}-2}\,\mathrm{sech}\big(2\lambda s\xi %
\big),  \notag
\end{eqnarray}%
where $0<s<1/2.$ Changing back to the $r$ variable we get
\begin{eqnarray}
\phi (r) &=&\,\mathrm{tanh}\big(2\lambda s\ln (r/r_{0})\big),  \notag \\%
[-0.2cm]
&&  \label{solring} \\[-0.2cm]
\chi (r) &=&\pm \sqrt{\frac{1}{s}-2}\,\mathrm{sech}\big(2\lambda s\ln
(r/r_{0})\big).  \notag
\end{eqnarray}%
which has a ring profile and $r_{0}$ can be identified with the ring radius
of the tube's cross section.

\begin{figure}[]
\hspace{-0.1cm}\includegraphics[{angle=0,width=4.35cm}]{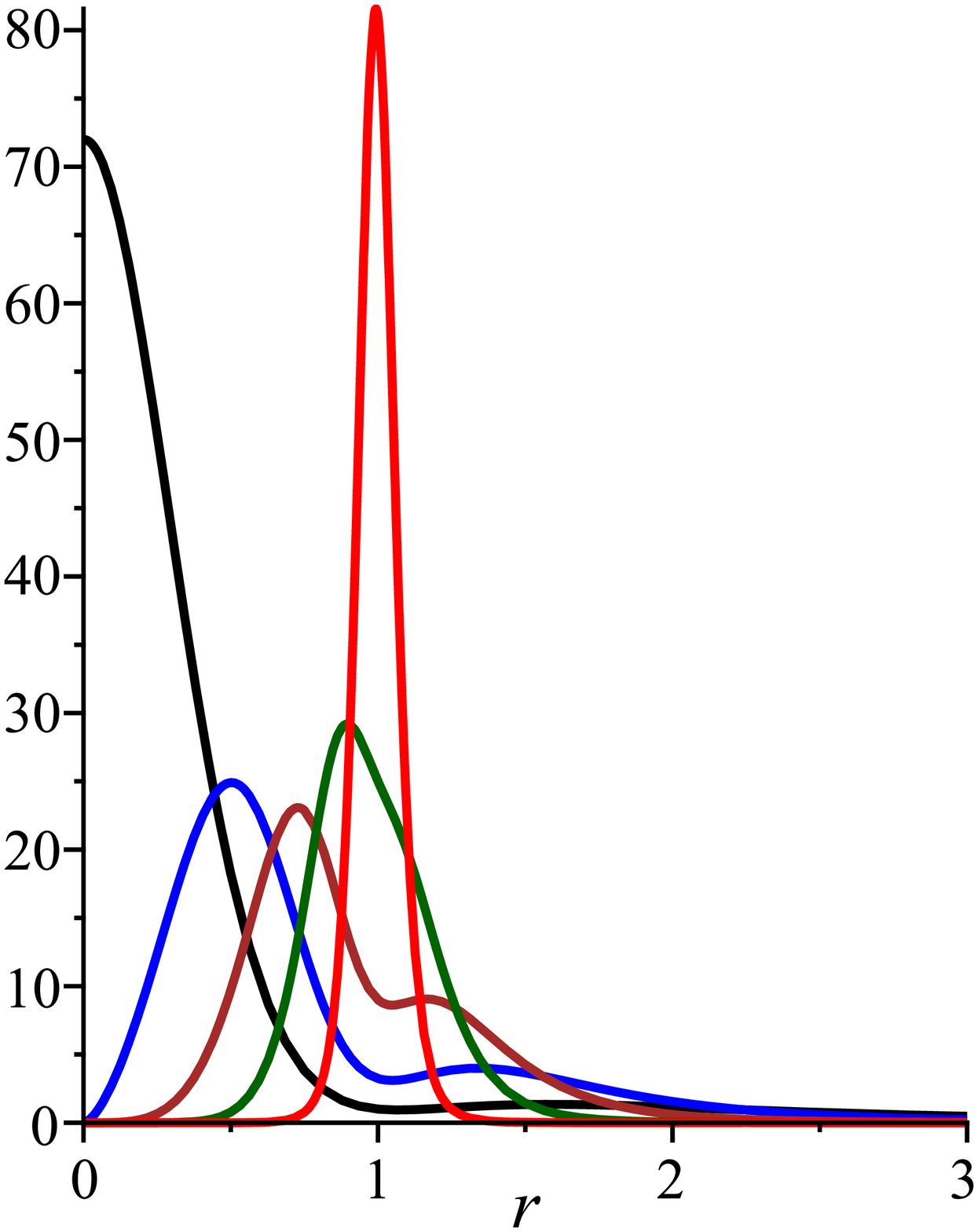}
\hspace{-0.2cm}\includegraphics[{angle=0,width=4.35cm}]{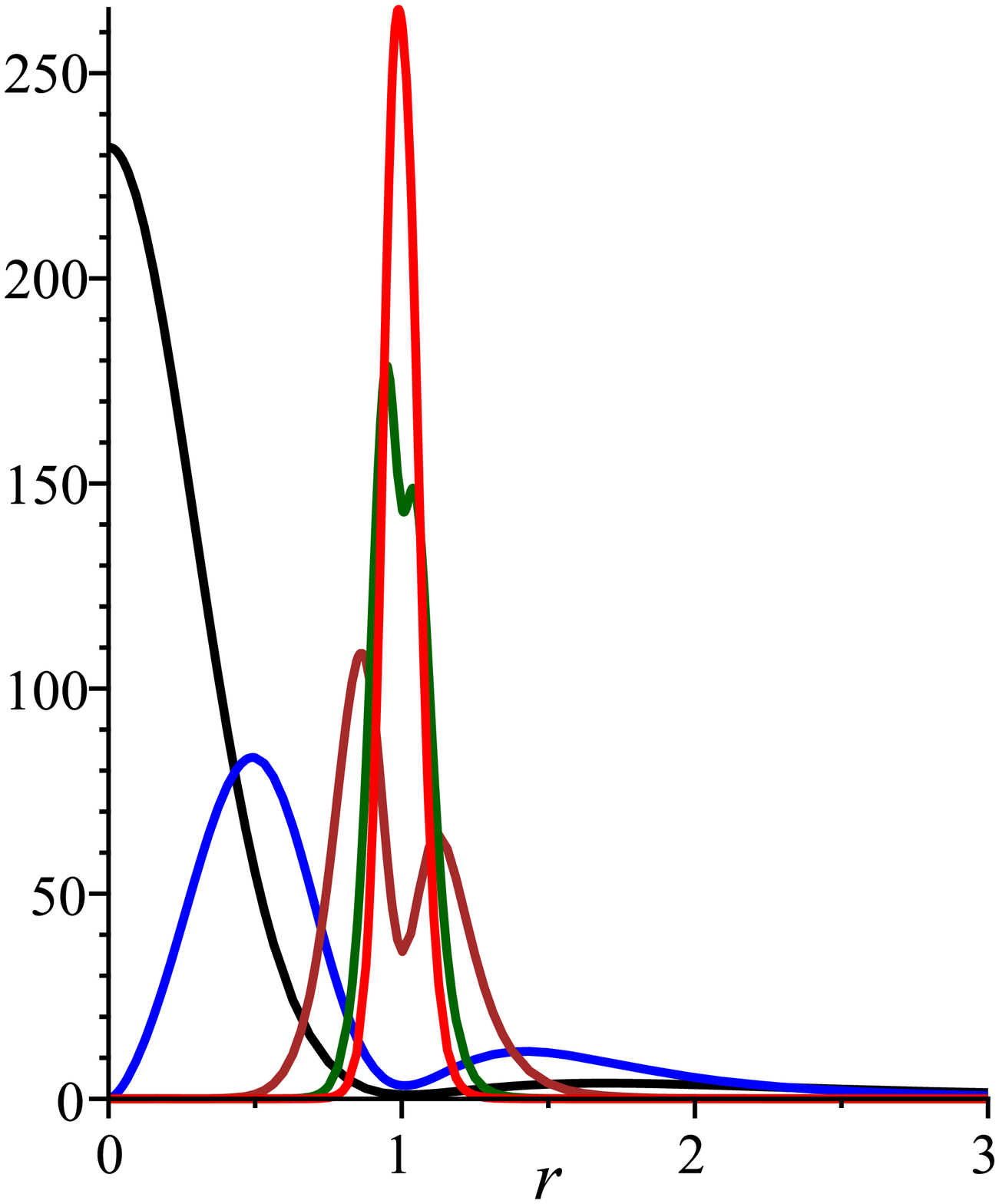}%
\vspace{-0.25cm}
\caption{Energy density $T^{00}(r)$ for $r_{0}=1$ : a) (left) $\protect%
\lambda =10$, $s=0.05$ (black), $s=0.09$ (blue), $s=0.15$ (brown), $s=0.25$
(dark green), $s=0.45$ (red); and b) (right) $\protect\lambda =30$, $s=1/60$
(black), $s=0.03$ (blue), $s=0.1$ (brown), $s=0.2$ (dark green), $s=0.27$
(red).}
\label{T00}
\end{figure}

The energy density of the $2$-dimensional defect is
\begin{eqnarray}
T_{00} &=&\frac{\left( 2\lambda s\right) ^{2}}{r^{2}}\mathrm{sech}^{4}\left[
2\lambda s\ln \left( \frac{r}{r_{0}}\right) \right]  \label{density} \\
&&~\times \left\{ 1+\left( \frac{1}{s}-2\right) \mathrm{sinh}^{2}\left[
2\lambda s\ln \left( \frac{r}{r_{0}}\right) \right] \right\}  \notag
\end{eqnarray}%
Here we consider $T_{00}$ finite in $r=0$, which restricts the parameters to
satisfy $\lambda s\geq \frac{1}{2}$ when $\lambda >1$. The Figs. \ref{T00}%
a-b depict the energy density $T_{00}\left( r\right) $ for fixed $r_{0}=1$
and several values of $\lambda $ and coupling constant $s$. We note that for
fixed $\lambda >1$ and $\frac{1}{2\lambda }\leqslant s<\frac{1}{2}$, the
behavior of the energy density changes from a lump centered in $r=0$ $\left(
s=\frac{1}{2\lambda }\right) $ to a peak centered around $r_{0}$\ $\left( s=%
\frac{1}{2}\right) $. For fixed $s$, the maximum amplitude of $T_{00}$
increases with $\lambda $, so large values of $\lambda $ produce more
interesting results. For large values of $\lambda $, there exists a value $%
s_{0}$ so that for $\frac{1}{2\lambda }<s<s_{0}$, the effects of the field $%
\chi $ are strong and the defect appears as a thick tube structure whose
center is localized between the origin and $r_{0}$. On the other hand, for $%
s_{0}\lesssim s<\frac{1}{2}$ it is clear the predominance of the field $\phi
$ and the defect looks like as a thin tube centered around $r_{0}$.

This means that for larger values of $\lambda $ we can characterize the
defect as a ring in the $2$-dimensional $yz$-plane, or as a cylindrical tube
in the $3$-dimensional space, oriented along the symmetry $x$ axis. The
influence of larger values of $s$ shows that the $\chi $ field is
responsible for the process of generating a thicker tube. The total energy
in the $yz$-plane is given by $E=8\pi \lambda /3$, which can be identified
with the mass of the ring, $M_{ring}$.

\section{fermion localization}

We are interested in the localization of (1,1)-dimensional fermions in an
infinite tube whose transversal section is the 2-dimensional ring. Here,
the tube in consideration is the one analyzed in the
previous section.  In considering fermionic states on a tube-like
defect, we must remark that the analysis about the existence or not of fermionic
zero modes was also considered in the other contexts. One can cite fermions in the field of an abelian and nonabelian \cite{string1,string2,string3} vortex solutions and more specifically neutrino zero modes on electroweak strings \cite{Zstring1, Zstring2, Zstring3}. 

In the following we consider the fermionic field coordinates as
$(x^{0},x^{1})=(t,x)$ and the ring coordinates as $(x^{2},x^{3})=(y,z)$.
Then, after neglecting the backreaction on the tube, we consider the following
fermionic action
\begin{equation}
S_{ferm}=\int \!\!dtdxdydz\!\!\left[ \frac{{}}{{}}\bar{\Psi}\Gamma
^{M}\partial _{M}\Psi -\eta F(\phi ,\chi )\bar{\Psi}\Psi \right] \!,
\label{action-ferm}
\end{equation}%
where the $\Gamma ^{0},\Gamma ^{1}\ $matrices are defined as
\begin{equation}
\Gamma ^{0}=i\gamma ^{0}=i\sigma ^{1},~\Gamma ^{1}=i\gamma ^{1}=\sigma ^{2},~
\label{matrices}
\end{equation}%
and $\Gamma ^{2},\Gamma ^{3}$\ are conveniently chosen to provide Schr\"{o}%
dinger's equations in the $yz-$plane whose potentials are supersymmetric
partners. Here $F(\phi (r),\chi (r))=F(r)$ is a function of the scalar
fields $\phi (r)$ and $\chi (r)$ giving the ring solution in Eqs. (\ref%
{solring}) and $\eta $ is the coupling constant.

After transforming the $yz-$plane to polar coordinates, the equation of
motion for $\Psi$ is found as
\begin{equation}
i\gamma ^{\mu }\partial _{\mu }\Psi +\left(
\begin{array}{cc}
\partial _{r}+\frac{i}{r}\partial _{\theta } & 0 \\
0 & -(\partial _{r}-\frac{i}{r}\partial _{\theta })%
\end{array}%
\right) \Psi -\eta F\Psi =0  \label{motion2}
\end{equation}%
where Greek letters $\mu ,\nu ...$ are for the $(t,x)$ coordinates and we
have chosen
\begin{equation}
\Gamma ^{r}=\sigma ^{3}=\left(
\begin{array}{cc}
1 & 0 \\
0 & -1%
\end{array}%
\right) ~\ ,~\ \ \Gamma ^{\theta }=i\mathbf{1=}\left(
\begin{array}{cc}
i & 0 \\
0 & i%
\end{array}%
\right) \,.  \label{matricesgamma}
\end{equation}

We decouple\ the coordinates $(t,x)$\ from $(r,\theta )$ by making the
decomposition
\begin{equation}
\Psi \left( t,x,y,z\right) =\sum_{n}\left(
\begin{array}{c}
R_{n}\left( r,\theta \right) \Psi _{Rn}\left( t,x\right) \\
L_{n}\left( r,\theta \right) \Psi _{Ln}\left( t,x\right)%
\end{array}%
\right) ,
\end{equation}%
and by imposing that $\Psi _{Rn}$ and $\Psi _{Ln}$\ are the chiral
components of a fermion satisfying the (1+1)-dimensional Dirac's equation%
\begin{equation}
\left( i\gamma ^{\mu }\partial _{\mu }-m\right) \left[
\begin{array}{c}
\Psi _{Rn} \\
\Psi _{Ln}%
\end{array}%
\right] =0\,,  \label{2d}
\end{equation}%
we can rewrite Eq. (\ref{motion2}) in the following set of equations for the
chiral amplitudes $L_{n}(r,\theta )$ and $R_{n}(r,\theta )$:
\begin{eqnarray}
\left( \partial _{r}-\frac{i}{r}\partial _{\theta }\right) L_{n}+\eta FL_{n}
&=&m_{n}R_{n}\,,  \label{eqacoplada1} \\
-\left( \partial _{r}+\frac{i}{r}\partial _{\theta }\right) R_{n}+\eta
FR_{n} &=&m_{n}L_{n}\,  \label{eqacoplada2}
\end{eqnarray}

Now we make the useful decomposition
\begin{eqnarray}
L_{n}(r,\theta ) &=&\sum_{\ell }\Lambda _{n\ell }(r)\,e^{i\ell \theta }
\label{decomposition_L} \\
R_{n}(r,\theta ) &=&\sum_{\ell }\varrho _{n\ell }(r)\,e^{i\ell \theta }
\label{decomposition_R}
\end{eqnarray}%
where $\ell \in Z$ and the functions $\Lambda _{n\ell }~,~\varrho _{n\ell }$
are finite in $r=0$.  Other decompositions are used in other contexts, see for instance, Refs. \cite{oda,gog,mello}

By combining the Eqs. (\ref{eqacoplada1}) and (\ref%
{eqacoplada2}) we attain the Schr\"{o}dinger-like equations for the scalar
modes $\Lambda _{n\ell }(r)$ and $\varrho _{n}(r)$%
\begin{eqnarray}
-\frac{d^{2}\Lambda _{n\ell }}{dr^{2}}+V_{sch}^{L}(r)\Lambda _{n\ell } &=&%
\hat{H}_{sch}^{L}\Lambda _{n\ell }=m_{n}^{2}\Lambda _{n\ell }\,,
\label{Eqsch_Left} \\
-\frac{d^{2}\varrho _{n\ell }}{dr^{2}}+V_{sch}^{R}(r)\varrho _{n\ell } &=&%
\hat{H}_{sch}^{R}\varrho _{n\ell }=m_{n}^{2}\varrho _{n\ell }
\label{Eqsch_Right}
\end{eqnarray}%
where the potentials are given by
\begin{eqnarray}
V_{sch}^{L}(r) &=&\frac{\ell \left( \ell +1\right) }{r^{2}}+2\eta \ell \frac{%
F}{r}-\eta \left( \partial _{r}F\right) +\eta ^{2}F^{2}\,,
\label{V_cal_left} \\
V_{sch}^{R}(r) &=&\frac{\ell \left( \ell -1\right) }{r^{2}}+2\eta \ell \frac{%
F}{r}+\eta (\partial _{r}F)+\eta ^{2}F^{2}\,.  \label{V_cal_right}
\end{eqnarray}
Then, we have transformed the equation for fermions in a set of independent
Schr\"{o}dinger-like equations for the amplitudes $\Lambda _{n\ell }$ and $%
\varrho _{n\ell }$, allowing to get our goal of finding massive modes and
analyzing their localization properties. The equations (\ref{Eqsch_Left})
and (\ref{Eqsch_Right}) allow us to adopt a probabilistic interpretation for
finding massive modes of both chiralities in the tube. Here we are mainly
interested in resonant states.

The Hamiltonians defining the Schr\"{o}dinger-like equations, (\ref%
{Eqsch_Left}) and (\ref{Eqsch_Right}), can be rewritten in terms of the
conjugate operators $\hat{A}$ and $\hat{A}^{^{\dag }}$\textbf{\ \ }%
\begin{equation}
\hat{A}=\frac{d}{dr}+\frac{\ell }{r}+\eta F~\ ,~\ \hat{A}^{^{\dag }}=-\frac{d%
}{dr}+\frac{\ell }{r}+\eta F  \label{sW}
\end{equation}%
as being $\ \hat{H}_{sch}^{L}=\hat{A}^{^{\dag }}\hat{A}$ and $\hat{H}%
_{sch}^{R}=\hat{A}\hat{A}^{^{\dag }}$, guaranteeing the eigenvalues $%
m_{n}^{2}$ to be nonnegative. In this way it is forbidden the existence of
tachyonic modes. The eigenfunctions $\Lambda _{n\ell }$ and $\varrho _{n\ell
}$ establish a complete set of orthonormal functions satisfying
\begin{equation}
\!\!\int \!\!dr\,\Lambda _{m\ell }\Lambda _{n\ell ^{\prime }}=\delta
_{mn}\delta _{\ell \ell ^{\prime }}=\!\!\int \!\!dr\,\varrho _{m\ell
}\varrho _{n\ell ^{\prime }}~,~\,\!\!\int \!\!dr\,\Lambda _{m\ell }\varrho
_{n\ell ^{\prime }}=0\,.  \label{orthono1}
\end{equation}
Further note that the action $S_{ferm}$ given by Eq. (\ref{action-ferm}) can
be integrated in the $(y,z)$ dimensions in order to obtain a standard action
for massive Dirac fermions
\begin{equation}
S_{ferm}=\sum_{n}\int dt\,dx\,{\bar{\Psi}_{n}[\gamma ^{\mu }\partial _{\mu
}-m_{n}]\Psi _{n}}  \label{action2d}
\end{equation}

Now we consider the issue of existence of the zero-mode $\chi _{0} $ which
is obtained from
\begin{equation}
\hat{A}\chi _{0}=0
\end{equation}%
For fixed $\ell $~the hamiltonian $H_{sch}^{R}$ is a quantum mechanical
supersymmetric partner of the hamiltonian $H_{sch}^{L}$ \ with
superpotential
\begin{equation}
\mathcal{W}=\frac{\ell }{r}+\eta F,  \label{SP}
\end{equation}
hence the solution for the zero mode is
\begin{equation}
\chi _{0}\propto r^{-\ell }\exp \left\{ -{\eta }\int_{0}^{r}dr^{\prime
}F\left( r^{\prime }\right) \right\} .
\end{equation}%
In our analysis we are considering the Yukawa coupling, $F(\phi ,\chi
)=\phi(r)\chi(r)$. Hence, because the integral $\displaystyle{%
\int_{0}^{r}\!\!\!dr^{\prime }F ( r^{\prime })}$ is finite for all $r$ the
zero mode in non-normalizable for all $\ell $. Since the zero-mode of $%
H_{sch}^{L} $ is non-normalizable, we conclude that the spectra of $%
H_{sch}^{L}$ and $H_{sch}^{R}$ are identical due to the spontaneous breaking of
supersymmetry in this quantum-mechanical system \cite{gan,ortiz}

\section{Numerical Results}

 For our purposes we
consider a simple Yukawa's coupling, $F(\phi )=\phi \chi $.
Interesting considerations for this and other couplings in models of two scalar fields can be found in \cite{cdh}.

\label{numerical} In order to investigate numerically the massive states,
firstly we consider the region near the origin (\textbf{\ }$r\ll r_{0}$)
where
\begin{equation}
F\left( r\right) \sim -2\sqrt{\frac{1}{s}-2}\biggl(\frac{r}{r_{0}}\biggr)%
^{2\lambda s}.
\end{equation}%
Then for $\lambda s\geq 1/2$, the functions $F/r$, $\partial _{r}F$ and $F^{2}$ are finite and the
potentials $V_{sch}^{L}(r)$ and$V_{sch}^{L}(r)$ are
dominated by the contributions of the angular momentum proportional to $%
1/r^{2}$. In this way, for left chirality the potential is
reduced to %
\begin{equation}
\tilde{V}_{sch}^{L}(r)\approx \frac{\ell \left( \ell +1\right) }{r^{2}},
\end{equation}%
and the solutions finite in $r=0$ are
\begin{eqnarray}
\Lambda _{n\ell }(r) &=&\sqrt{r}J_{\ell +\frac{1}{2}}(m_{n}r)~,\ \ \ell \geq
1  \label{n4a} \\
\Lambda _{n\ell }\left( r\right) &=&\sqrt{r}Y_{\ell +\frac{1}{2}}\left(
m_{n}r\right) \ ,\ \ell \leq -2,  \label{n4b} \\
\Lambda _{n\ell }(r) &=&\sqrt{r}Y_{\frac{1}{2}}(m_{n}r)~,\ \ \ell =0,
\label{n4c} \\
\Lambda _{n\ell }(r) &=&\sqrt{r}J_{-\frac{1}{2}}(m_{n}r)~,\ \ \ell =-1.
\label{n4d}
\end{eqnarray}%
For right chirality we have
\begin{equation}
\tilde{V}_{sch}^{R}(r)\approx \frac{\ell ^{2}-\ell }{r^{2}},
\end{equation}%
whose solutions finite in $r=0$  are
\begin{eqnarray}
\varrho _{n\ell }(r) &=&\sqrt{r}Y_{\ell -\frac{1}{2}}(m_{n}r)~,\ \ \ell \leq
-1  \label{n5a} \\
\varrho _{n\ell }\left( r\right) &=&\sqrt{r}J_{\ell -\frac{1}{2}}\left(
m_{n}r\right) \ ,\ \ell \geq 2,  \label{n5b} \\
\varrho _{n\ell }(r) &=&\sqrt{r}J_{-\frac{1}{2}}(m_{n}r)~,\ \ \ell =0,
\label{n5c} \\
\varrho _{n\ell }(r) &=&\sqrt{r}Y_{\frac{1}{2}}(m_{n}r)~,\ \ \ell =1.
\label{n5d}
\end{eqnarray}

Hence, for each value of $\ell $, Eqs. (\ref{n4a})-(\ref{n4d}) or (\ref{n5a}%
)-(\ref{n5d}) are used as an input for the Runge-Kutta-Fehlberg method that
produces a fifth order accurate solution.

We now define the probability for finding fermions inside the tube of radius $r_0$ as
\begin{equation}
P_{tube}={\frac{\displaystyle\int_{r_{min}}^{r_{0}}\!\!\!\!dr\,|\Psi
_{n}(r)|^{2}}{\displaystyle\int_{r_{min}}^{r_{max}}\!\!\!\!dr\,|\Psi
_{n}(r)|^{2}}}\,.  \label{n6}
\end{equation}%
Here $r_{min}\ll r_{0}$ is used as the initial condition and $r_{max}$ is
the characteristic box length used for the normalization procedure, being a
value where the Schr\"{o}dinger potentials are close to zero and where the
massive modes oscillate as plane waves.

\begin{figure}[]
\hspace{-0.2cm}%
\includegraphics[{angle=0,width=4.3cm}]{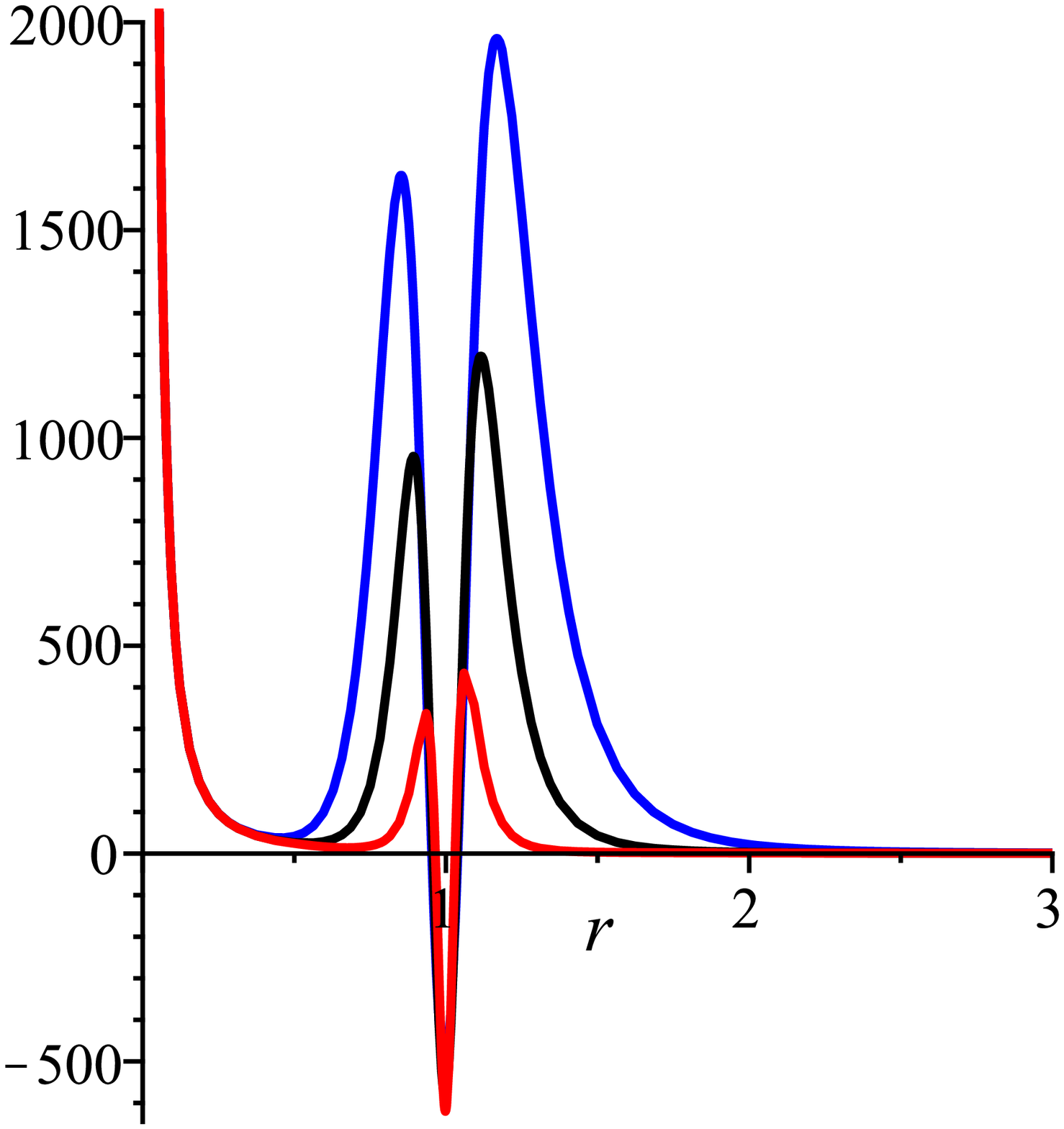}
\hspace{-0.2cm} %
\includegraphics[{angle=0,width=4.3cm}]{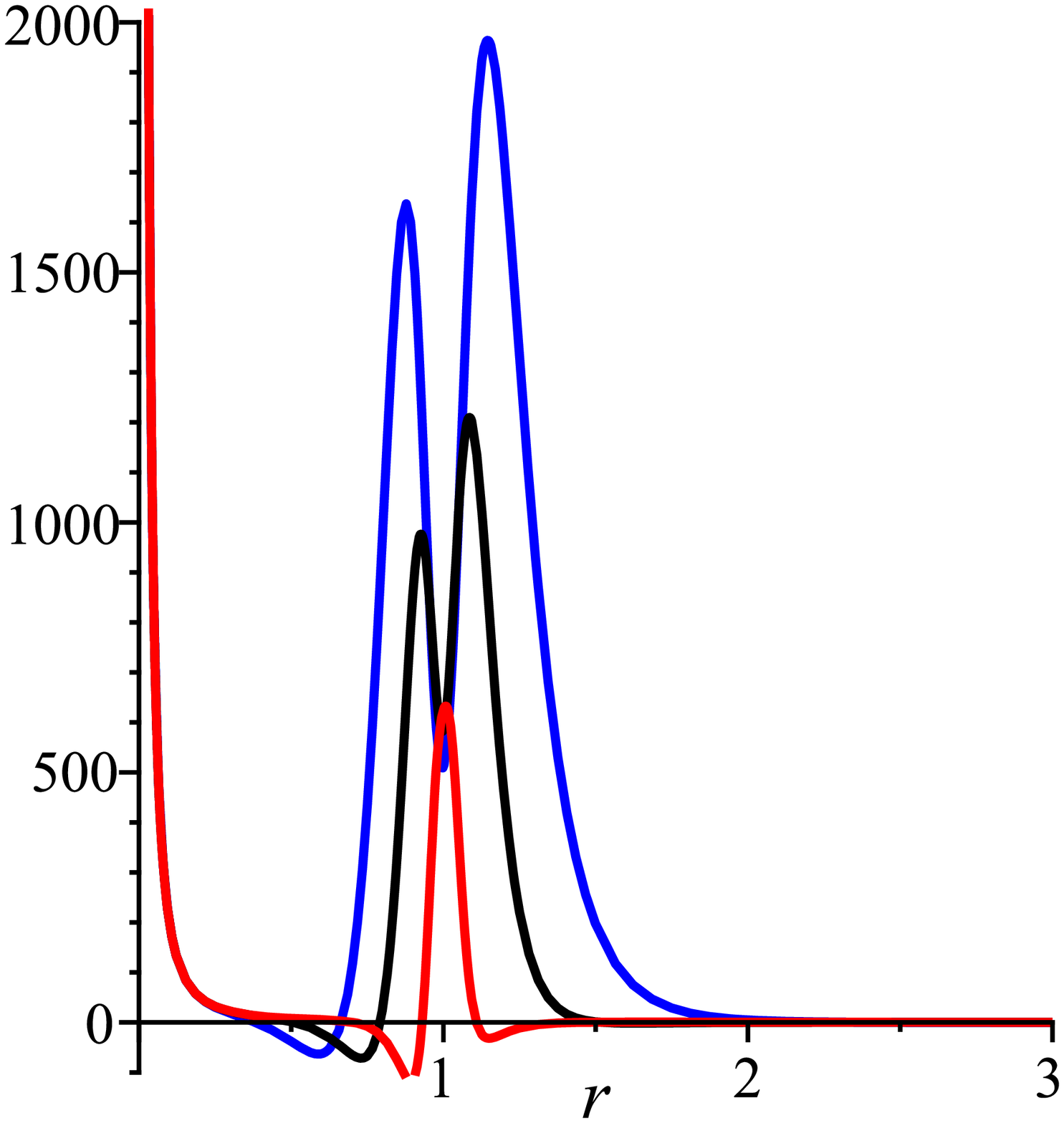} 
\hspace{0.2cm}%
\includegraphics[{angle=0,width=4.3cm}]{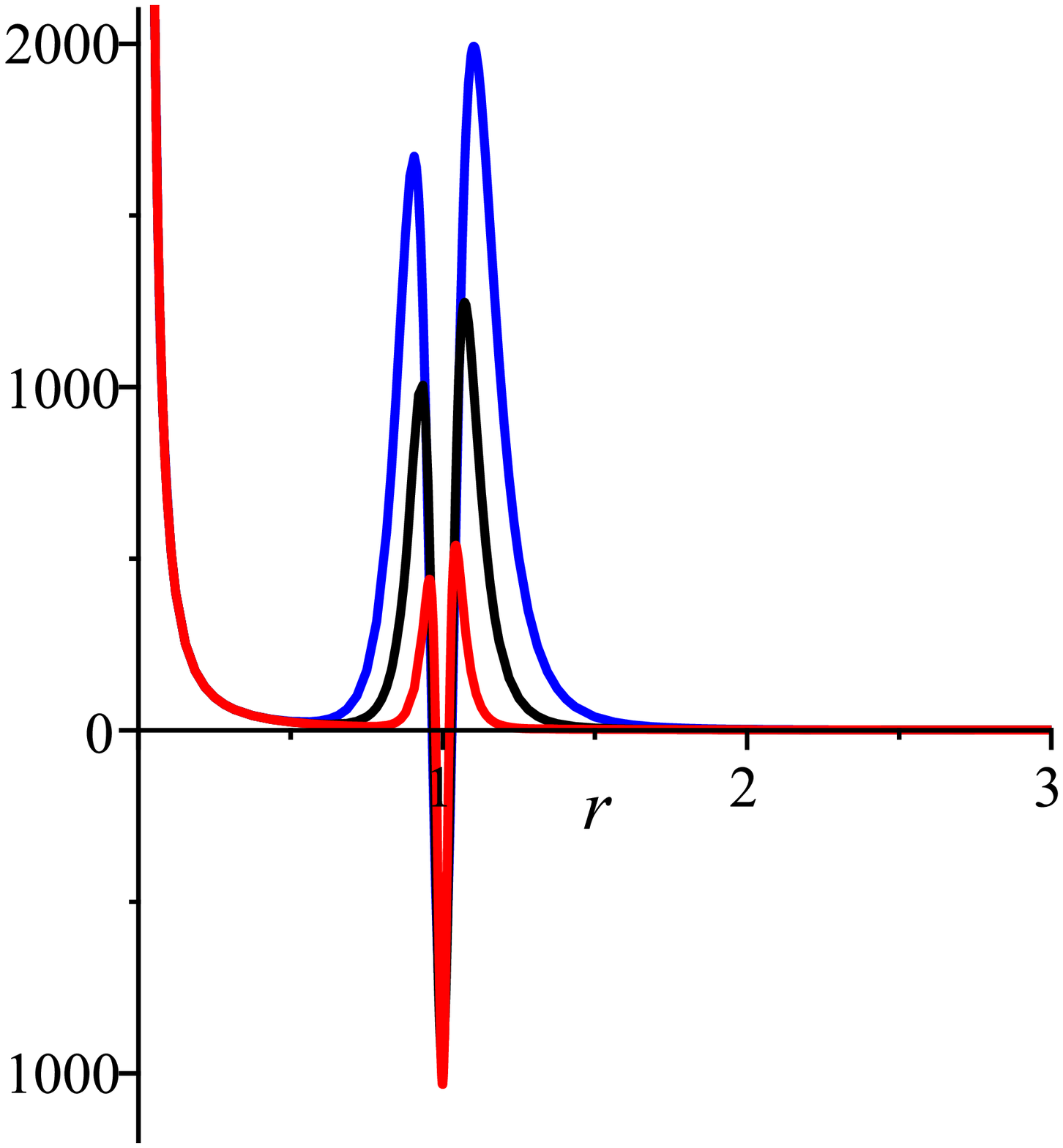}
\hspace{-0.2cm} %
\includegraphics[{angle=0,width=4.3cm}]{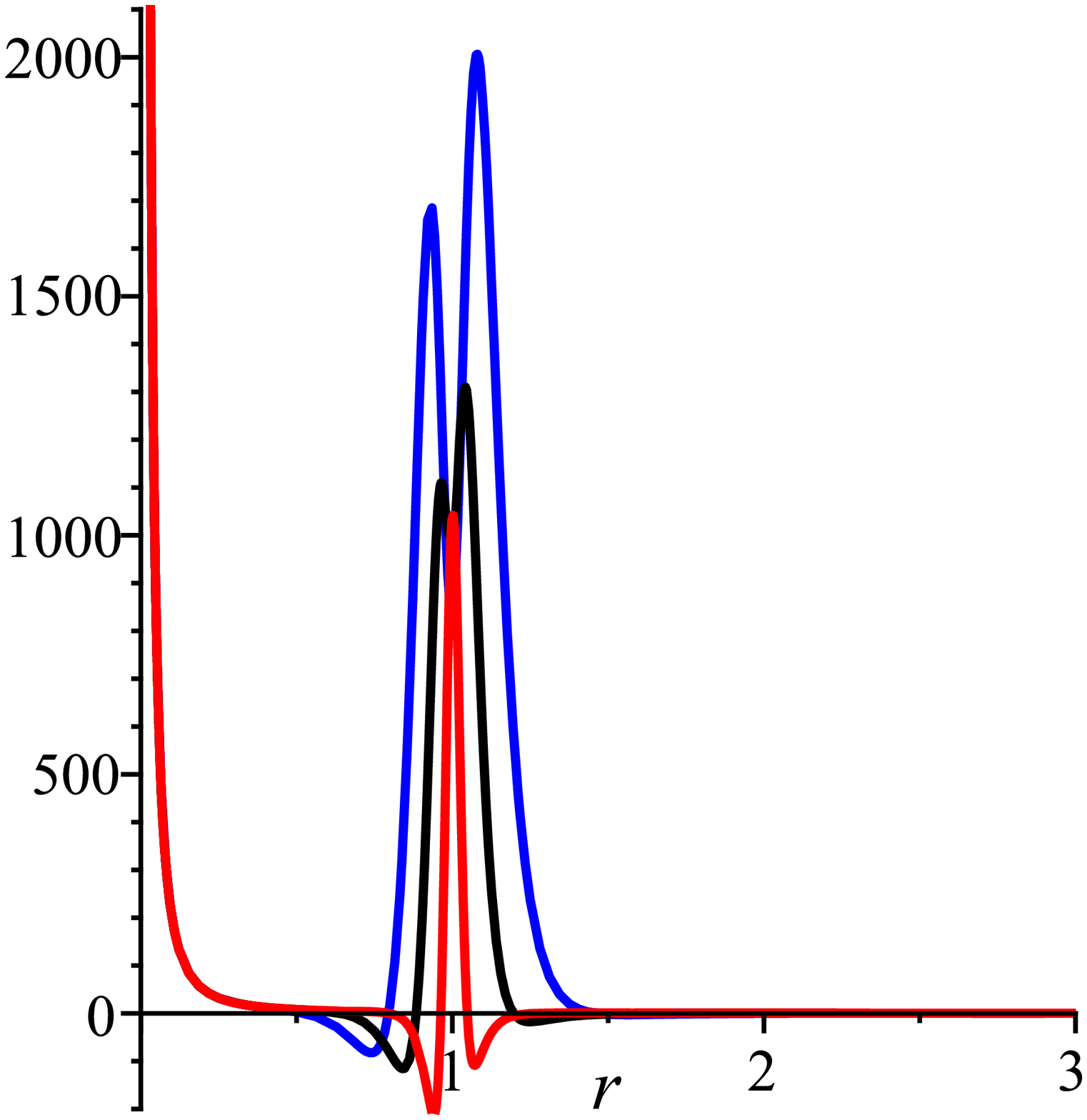}
\vspace{-0.15cm}
\caption{Schr\"{o}dinger-like potentials $V_{sch}^{L}$ (left) and $%
V_{sch}^{R}$ (right) for $\ell =2$. We fix $r_0=1$, and $\protect\eta=30$.
We have a) $\protect\lambda=30$ (upper figures) and b) $\protect\lambda=50$
(lower figures). In all figures $s=0.1$(blue), $s=0.15$(black), $s=0.3$%
(red). }
\label{potVsch_l}
\end{figure}

From the energy density considerations after Figs. \ref{T00}a-%
\ref{T00}b, larger values of $\lambda $ favor the existence of a
Schr\"{o}dinger potential with structure similar a tube\ barrier em $r=r_{0}
$. \ The Fig. \ref{potVsch_l} depicts the Schr\"{o}dinger-like
potential $V_{sch}^{L}(r)$ and $V_{sch}^{R}(r)$ for $%
\ell =2$, $\lambda =30$, $50$, and fixed $\eta =30$
and $r_{0}=1$. The potentials in general diverge in $r\rightarrow 0$%
, assume a form of a barrier around $r=r_{0}$ and
asymptote to zero as $r\rightarrow \infty $, indicating the
possible presence of resonances. The increasing of $\eta $ turns
the barrier of the potential higher, whereas the increasing of $\lambda $
 turns it thinner. We noted that $\ell $ influences on
the behavior of the potential for $r<r_{0}$ but has no sensible
influence on the barrier and, we also observe that the increasing of $r_{0}$
 turns the potential barrier wider.

\begin{figure}[]
\includegraphics[{angle=0,width=4cm}]{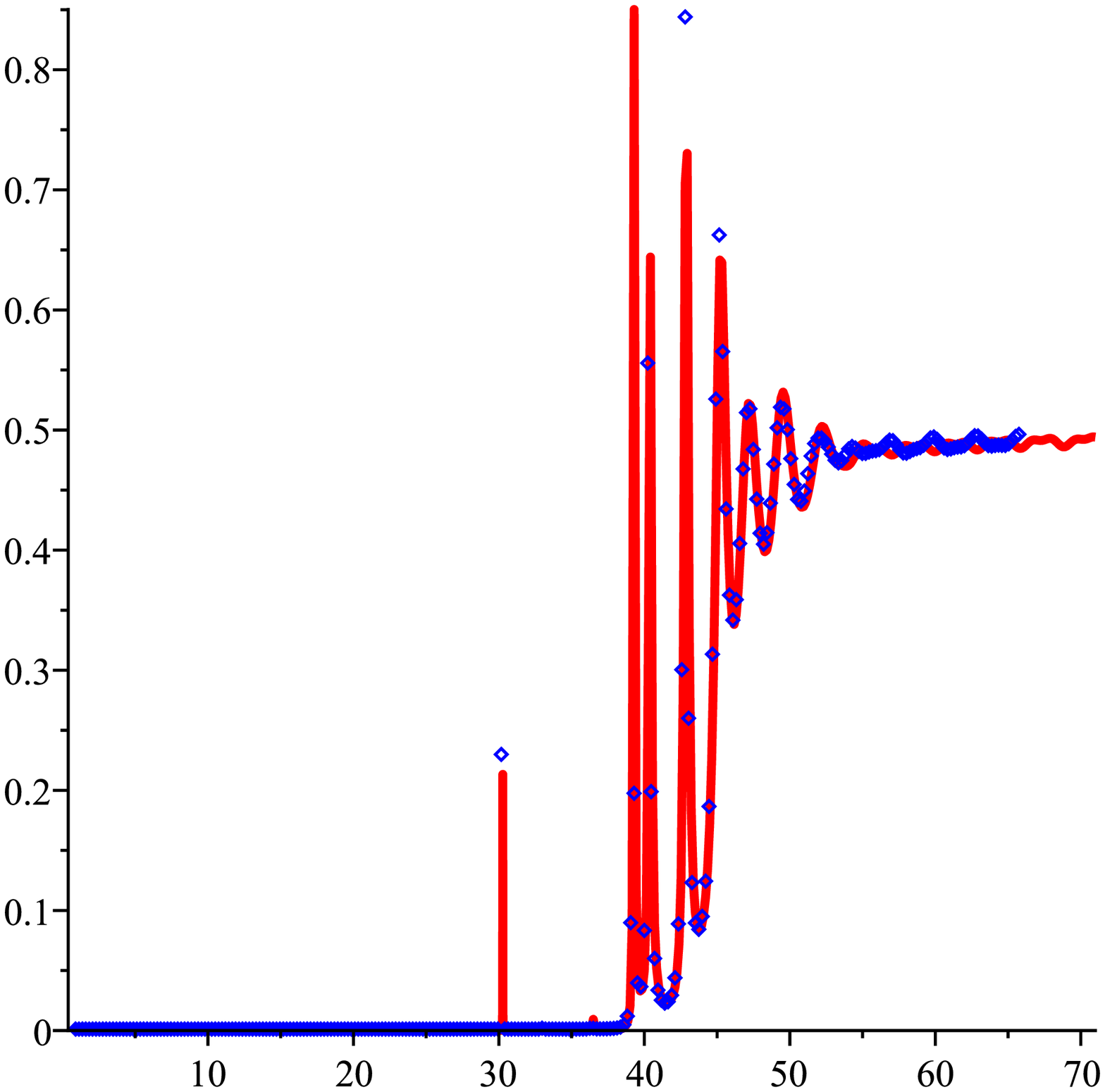} 
\includegraphics[{angle=0,width=4cm}]{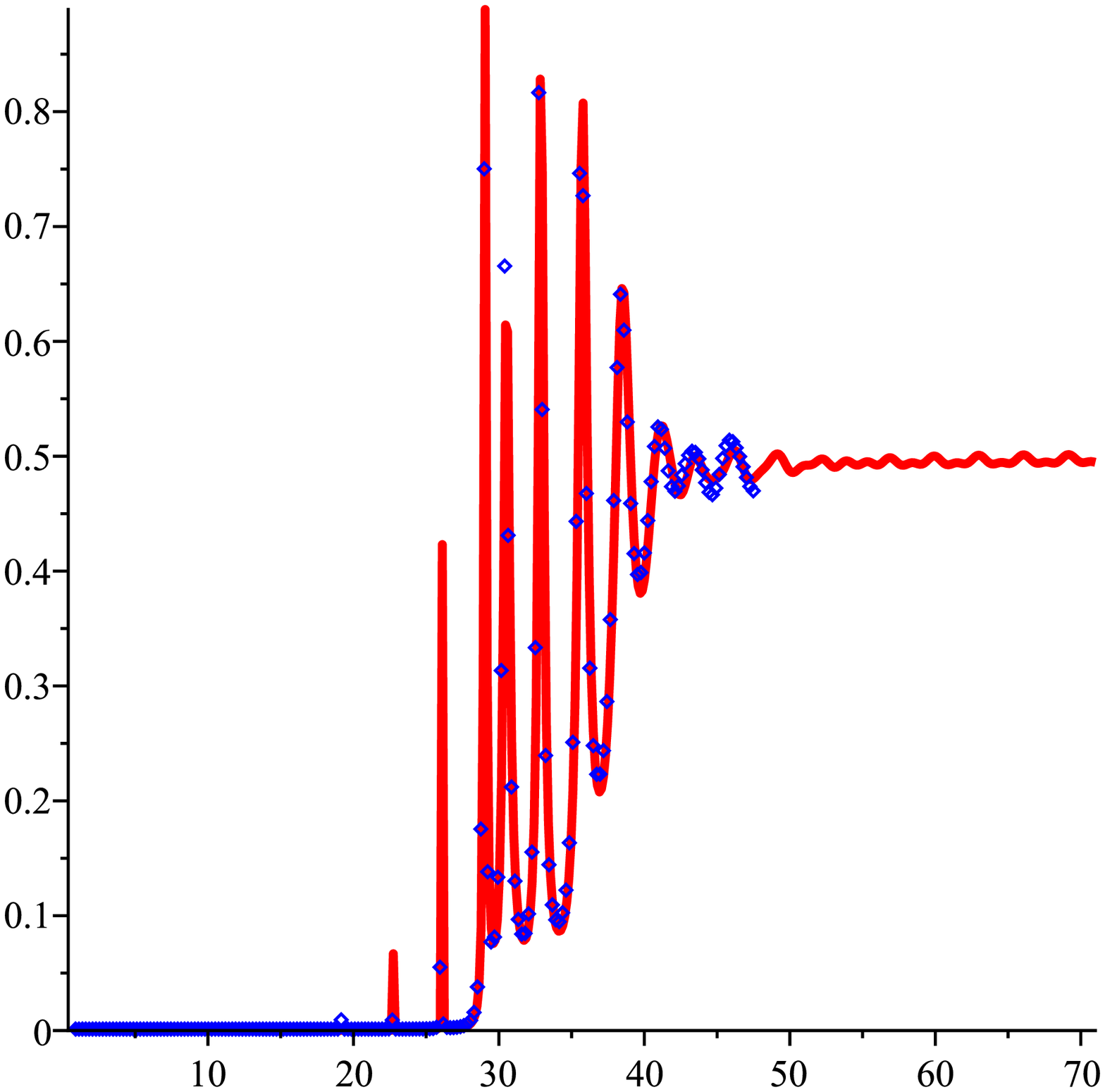} 
\includegraphics[{angle=0,width=4cm}]{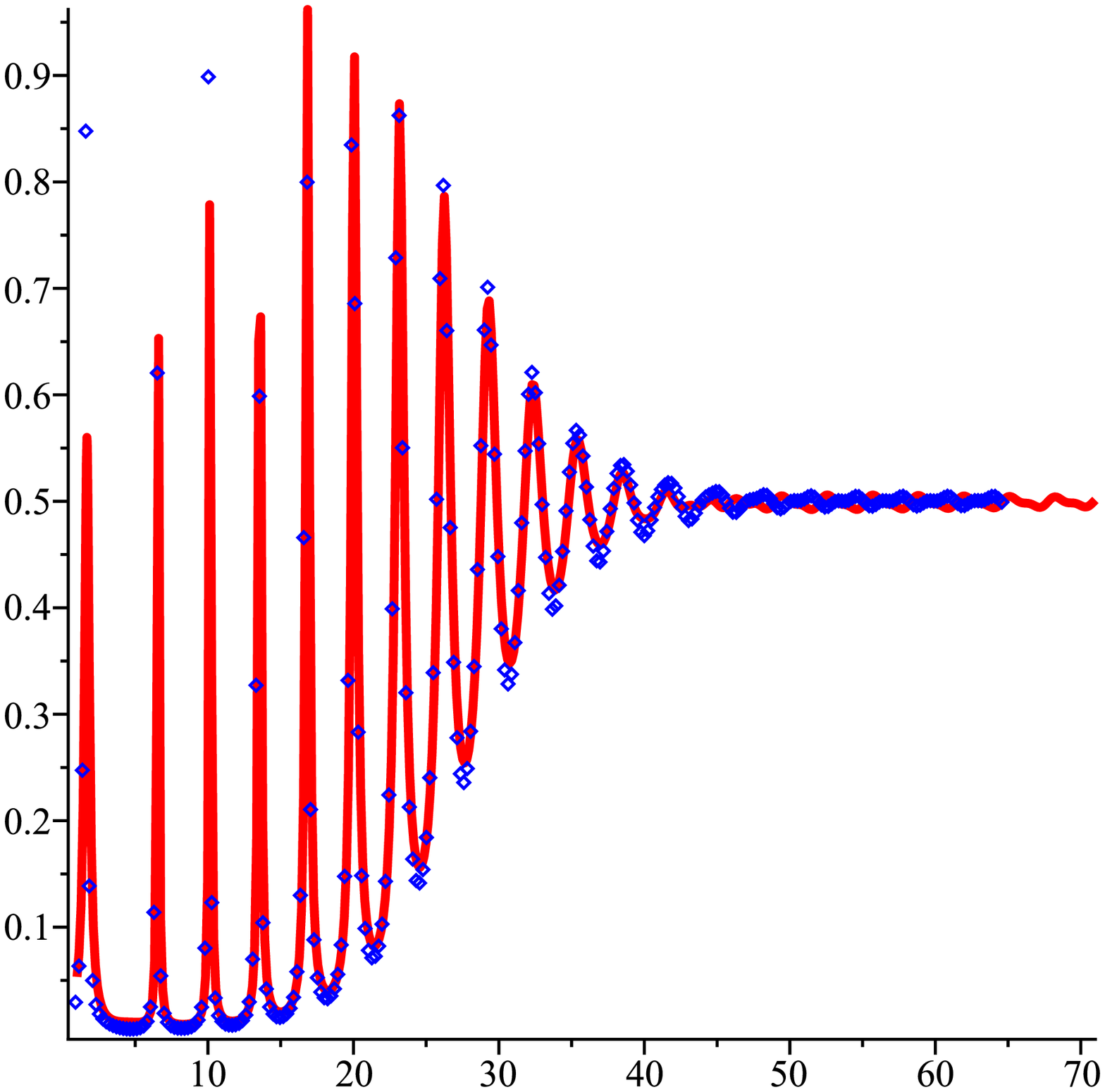} 
\caption{$P_{tube}(r)$ for left-(red line) and right-(dots) chirality
fermions with $\ell =2$, $r_0=1$ $\protect\eta=30$, $\protect\lambda=30$, a)
$s=0.10$, b) $s=0.15$, c) $s=0.30$.}
\label{Ptuberight}
\end{figure}

Figs. \ref{Ptuberight}a-c show some results of $P_{tube}(r)$
\ for left- and right-fermions. The plots are for $\ell =2$, $
r_{0}=1$, $\eta =30$ and $\lambda =30$ and for
various values of $s$. The plots shows several thin peaks of
resonances. The thinner is the peak, the larger is the lifetime of the
corresponding resonance. We note for $s\sim 0.10$ there are fewer
but longer-lived resonance peaks, in comparison to $s\sim 0.30$.  Note also that
in general the masses of the resonances $m_{ress}$ increase gradually for lower values of $s$, but the relation $M_{ring}\ll m_{ress}$ was always verified,
guaranteeing a condition for no backreaction of
the fermions in the ring.
The larger lifetime of the modes agrees with the correspondingly larger
barrier of the Schr\"{o}dinger-like potential around $r=r_{0}$
(compare with Figs. \ref{potVsch_l} ). This property of $V_{sch}$ is also
responsible for the presence of resonance modes, in general, more massive
for lower values of $s$ $\left( s\rightarrow \frac{1}{2\lambda }%
\right) $ which\ are closely related to the larger influence of the
$\chi $ field in the internal structure of the defect. However, in
this limit the number of the resonances is reduced because the mass values
can not be higher than the tube-barrier.\ In this way, for finding
long-lived resonances, there appears to be a physical compromise between a
thinner tube (larger values of $s$ which unfavor the presence of the $
\chi $ field) and the Yukawa coupling $\phi\chi$ (smaller values of $s$%
, related to a greater influence of the field $\chi $).

\section{Remarks and conclusions}

We have studied the localization of (1+1)-dimensional fermionic
fields in a generalized tube-like topological defect whose cross-section is
a ring constructed\ with two scalar fields. Firstly, we have considered a
general coupling between the defect and the fermion field  carefully
contructed to provide a supersymmetric quantum mechanical description of
the\ chiral amplitudes related to the left- and right-fermionic components. \
Consequently, the Hamiltonians describing the chiral amplitudes\  are
supersymmetric partners\ forbiding the existence of tachyonic modes. For the Yukawa coupling
 $F(\phi,\chi)=\eta \phi \chi$
 it was shown that the zero-mode is non-normalizable and that the spectra
of both chiralities are identical due to the spontaneous breaking of
supersymmetry. Such result is corroborate by the numerical analysis of the
Hamiltonian spectra. \ Also, it was found that larger couplings $\eta $ and $\lambda $ are more effective for finding resonances
after a fine-tuning of the contant $s$ characterizing the internal
structure of the defect.

As a further comment we would like to point out that the Yukawa coupling is
useful for studies of the electromagnetic charge of the ring and some
effects like charge fractionalization. These studies are currently under
consideration.

\section*{Acknowledgements}

The authors thank CAPES, CNPq and FAPEMA for financial support. R Casana and AR Gomes
thank MM Ferreira Jr. for discussions.

\end{document}